\documentstyle[prl,aps]{revtex}
\input epsf
\newcommand{\be}{\begin{equation}}
\newcommand{\ee}{\end{equation}}

\begin{document}

\preprint{LAUR 96-2036}

\draft

\twocolumn[

\title{Density of kinks after a quench: \\
       When symmetry breaks, how big are the pieces?}

\author{Pablo Laguna${}^{(1,2)}$ and Wojciech Hubert Zurek${}^{(1)}$}
 
\address{
${}^{(1)}$ Theoretical Astrophysics, T-6, MS B288 \\
Los Alamos National Laboratory, Los Alamos, NM 87545, USA\\
${}^{(2)}$ Department of Astronomy \& Astrophysics and \\
Center for Gravitational Physics \& Geometry\\            
Penn State University, University Park, PA 16802, USA}

\date{\today}
\maketitle

]

\begin{abstract}
Numerical study of order parameter dynamics in the course 
of second order (Landau-Ginzburg) symmetry breaking transitions shows that 
the density of topological defects, kinks, is proportional to
the fourth root of the rate of the quench. This confirms the more general theory
of domain-size evolution in the course of symmetry breaking transformations
proposed by one of us \cite{Zurek85}. Using these ideas, it is possible to compute
the density of topological defects from the quench timescale and
from the equilibrium scaling of the correlation length
and relaxation time near the critical point. 

\end{abstract}


The dynamics of symmetry breaking phase transitions has been the focus 
of attention because of its implications for cosmological 
scenarios\cite{Zurek85,Zeldovich,Kibble,Vilenkin,Davis} and its importance in 
the context of condensed matter 
physics\cite{Chuang,Hendry,Bray96,Ruutu,Bauerle,Zurek96}. The likely cosmological
setting for this process involves a second order phase transformation
associated with a non-conserved order parameter $\varphi$. Initially, the system
is near equilibrium in the symmetric phase with $\langle\varphi\rangle = 0$.
When the effective potential assumes the ``sombrero'' shape, $\varphi$
is forced to make a choice of one of many possible broken symmetry true vacua. 
As pointed out by Kibble\cite{Kibble} in the cosmological context,
the finiteness of the speed of light implies that the choices of 
the new broken symmetry vacuum must be independent in sufficiently distant 
regions of space. Therefore, when the homotopy group, characterizing the
relation between the manifold of broken symmetry vacua and the space in 
which $\varphi$ evolves, is non-trivial, locally independent choices 
of the new vacuum will lead to topologically stable objects such as monopoles,
cosmic strings, or domain walls. The initial density of such 
defects is of great interest and cannot be deduced from the scenario sketched 
above, save for a much too generous lower limit which follows from 
the lightcone causality alone \cite{Ginzburg}. 

\begin{figure}[h]
\leavevmode
\epsfxsize=3.2truein\epsfbox{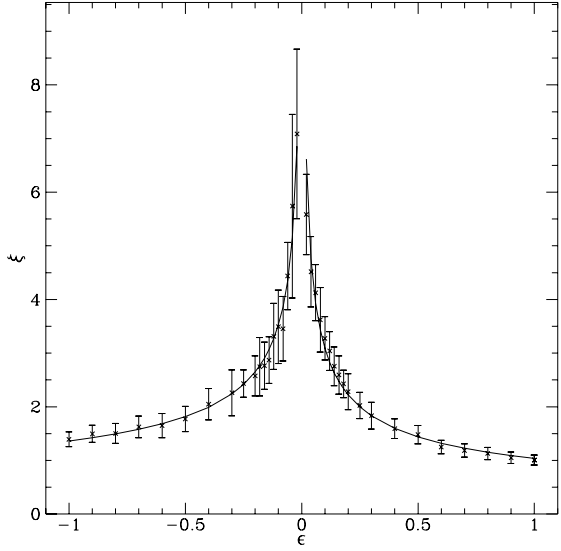}
\caption[figure1]{\label{figure1}
Characteristic equilibrium correlation (or healing) length,  
$\xi = \xi_0 / | \epsilon |^{\nu},$
for the system under investigation. The best fitting 
yields $\xi_0 = 1.38 \pm 0.06$, $\nu = 0.41 \pm 0.03$ ($\chi^2 = 1.2$) above $T_C$,
and  $\xi_0 = 1.02 \pm 0.04$, $\nu = 0.48 \pm 0.02$ ($\chi^2 = 3.7$) below $T_C$,
close to the Landau-Ginzburg exponent of $\nu = 1/2$.}
\end{figure}

As suggested in \cite{Zurek85}, it is also possible to study ``cosmological''
mechanisms of defect formation in condensed matter.
In this context, the estimate of defect density based on speed of light
arguments is considerably less useful than in cosmological situations. 
Thus, along with the proposal for condensed matter tests of defect formation
in cosmological scenarios, a theory aimed at computing the defect density was 
developed\cite{Zurek85,Zurek96}. The key idea is to realize 
that the order parameter can only become correlated through
its dynamics. Furthermore, in the vicinity of the critical temperature
$T_C$, order parameters exhibit universal behavior characterized by two simultaneously
occurring divergences. That is, when parametrized in terms of the
relative temperature $ \epsilon = (T_C - T)/T_C  $, both
the equilibrium {\it healing length} (also known as the
{\it correlation length}) 
\be
\xi ~ = ~ \xi_0 / | \epsilon |^{\nu}   
\label{eq:xi}
\ee
and the {\it dynamical relaxation time}  
\be
\tau ~ = ~ \tau_0 / | \epsilon |^{\mu} 
\label{eq:tau}
\ee
will simultaneously diverge at the critical temperature (see Fig.\ \ref{figure1}).
Above, $\xi_0$ and $\tau_0$ 
characterize the low temperature ($T=0, \ \epsilon = 1$) values of the healing 
length and relaxation time, respectively. 

The healing length and the dynamical relaxation time have similar
physical significance. 
$\xi$ is the distance over which the order parameter returns to its 
equilibrium value when perturbed, for instance, by the boundary conditions. 
$\xi$ is also the typical scale of the perturbations; that is, to the leading 
order, the correlation function of perturbations away from equilibrium value,
$\delta \varphi(x,t)$, behaves as
\be
\langle \delta \varphi(x,t), \delta \varphi(x + \Delta,t) 
\rangle \ \sim \ \exp(-|\Delta|/\xi) \ .
\label{eq:perturb}
\ee 
Below $T_C$, the spatial order is established on scales much larger 
than $\xi$; however, the scale over which the healing occurs ---for example, from
the ``wounds'' inflicted by topological defects--- is characterized 
by $\xi$. 
Similarly, $\tau$ characterizes the time required for the order parameter 
to relax to its equilibrium value. During the phase transition,
in the immediate vicinity of the critical temperature, the motion will
be often overdamped (that is, dominated by the first time derivative of the order
parameter) and $\tau = \tau_0 / |\epsilon|$.

The estimate of the density of defects put forward 
in Ref.\cite{Zurek85} is based on a linear quench, 
\be
\epsilon \ = \ t/\tau_Q \ ,
\label{eq:epsilon}
\ee
which is expected to be a suitable approximation in the neighborhood of $T_C$.
In (\ref{eq:epsilon}), $\tau_Q$ is the {\it quench timescale} and $t$ the time before ($t<0$) 
and after ($t>0$) the transition ($\epsilon = 0$). 
When the critical temperature is crossed at the fixed rate given by 
Eq.~(\ref{eq:epsilon}), there will come a moment when the order parameter 
---because of the critical slowing down implied by Eq.~(\ref{eq:tau})--- 
will be simply unable to adjust its
actual correlation length to the equilibrium $\xi$ given by Eq.~(\ref{eq:xi}). 
This will occur when the time remaining until the transition, $\hat t$, equals 
the relevant dynamical relaxation time $\tau(\hat t)$, that is, when
\be
\tau(\hat t) = \tau_0/(|\hat t|/\tau_Q)^{\mu} = \hat t \ . 
\label{eq:that}
\ee
This equation yields a relative temperature
\be
\hat \epsilon = \hat t / \tau_Q = (\tau_0 / \tau_Q)^{1 \over 1 + \mu},
\label{eq:ehat}
\ee
at which the evolution of the order parameter will cease to follow equilibrium.
At this point, the correlation length of $\varphi$ will cease to diverge 
in accord with Eq.~(\ref{eq:xi}), 
as the phase transition region is
traversed. Instead, $\xi$ will reach a value approximately given by
\be 
\hat \xi = \xi_0 / |\hat \epsilon |^{\nu} 
= \xi_0 (\tau_Q/\tau_0)^{\nu \over 1+\mu} \ .
\label{eq:xihat}
\ee 
Regions more distant than $\hat \xi$ will be forced to select the new vacuum 
independently.

An essentially equivalent estimate of the characteristic domain size is
obtained from the ``causal horizon'' idea, but only when the appropriate velocity of
the signal propagation on $\varphi$ is adopted\cite{Zurek85,Zurek96}.
That velocity is of the order of $v = \xi / \tau$ for the dynamical processes
correlating different regions of the order parameter.
Thus, the size of the corresponding causal horizon size (and, therefore, 
the size of the domain) will be no more than 
$h(\hat t) = v(\hat t) \hat t \simeq \hat \xi$. 

In any case, the size of independently selected domains of the new vacuum
will be given, to the leading order, by $\hat \xi$, which will also determine 
the typical separation of topological defects and, therefore, the initial 
defect density $n$. For instance, for monopoles 
$ n \approx 1 / (f \hat \xi)^D \ , $
with $D$ the space dimension and $f \sim {\cal O}(1)$ 
a factor presumably somewhat larger than unity. $f$ takes into account 
the possibility that the independent choices of the vacuum may be similar anyway,
and that the correlations length will slowly 
grow following the instant $\hat t$ due to diffusion, etc.; the key prediction 
of the theory of \cite{Zurek85} is in any case the {\it scaling}, as 
is usually the case in the critical phenomena.

The aim of the work in this Letter is to test this theory with 
a computer experiment.  The immediate motivation of our research comes from 
recent superfluid experiments\cite{Hendry,Ruutu,Bauerle}. These experiments
follow the original suggestion in Ref. \cite{Zurek85} and appear to support the estimate 
of defect formation based on $\hat \xi$. While the quench-generated density 
of vortex lines is somewhat uncertain, it is, nonetheless, in accord with the 
theory summed up above but in conflict with the ideas based on activation
and Ginzburg temperature\cite{Ginzburg}. However, these experiments
were, at least so far, unable to vary the quench time $\tau_Q$. 
Thus, it was not possible to systematically test the key predictions of Ref. 
\cite{Zurek85}, namely the power law dependence of the size of the fragments
of the broken symmetry vacuum on the quench rate $\tau_Q^{-1}$ given by Eq.~(\ref{eq:xihat}), 
and the complementary dependence of the initial number of defects.

To investigate this issue, we considered the numerical evolution 
of a one-dimensional system \cite{Note} for a real field $\varphi$
according to the equation of
motion derived from the Landau-Ginzburg potential
$
V(\varphi) = (\varphi^4 - 2 \epsilon \varphi^2 + 1)/8 \ .
$
The system is assumed to be in contact with
a thermal reservoir. Thus, it obeys the Langevin equation
\be
\ddot \varphi + \eta \dot \varphi - 
\partial_{xx} \varphi + \partial_{\varphi} V(\varphi) = \vartheta \ . 
\label{eq:motion}
\ee
The noise term $\vartheta$ has correlation properties
\be
\langle \vartheta (x, t), \vartheta (x', t') \rangle =  2 \eta \theta \delta(x'-x)
\delta(t'-t) \ . 
\label{eq:noise}
\ee
In Eq.~(\ref{eq:noise}), $\eta$ is the overall damping constant which also helps
characterize the amplitude of the noise through Eq.~(\ref{eq:motion}). The parameter
$\epsilon$ measures the distance from the phase transition 
and varies according to Eq.~(\ref{eq:epsilon}); that is,
$\epsilon = \hbox{min}(1,t/\tau_Q)$.
$\theta$ describes the temperature of 
the reservoir and is kept constant throughout this work. 
This separate parametrization of $\epsilon$ and $\theta$
was adopted to correspond to the situation in superfluid
He$^4$, where the symmetry breaking can be induced by the change of the pressure
and occurs with inconsequential adjustments of the absolute temperature\cite{Zurek85,Hendry}.
The second order time evolution allows us in principle to make contact with cosmology, 
but, in the regime considered here, the evolution is dominated by the dissipative 
term $\eta \dot \varphi$. Hence, in effect, we are dealing with the time
dependent Landau-Ginzburg equation.

\begin{figure}[h]
\leavevmode
\epsfxsize=3.2truein\epsfbox{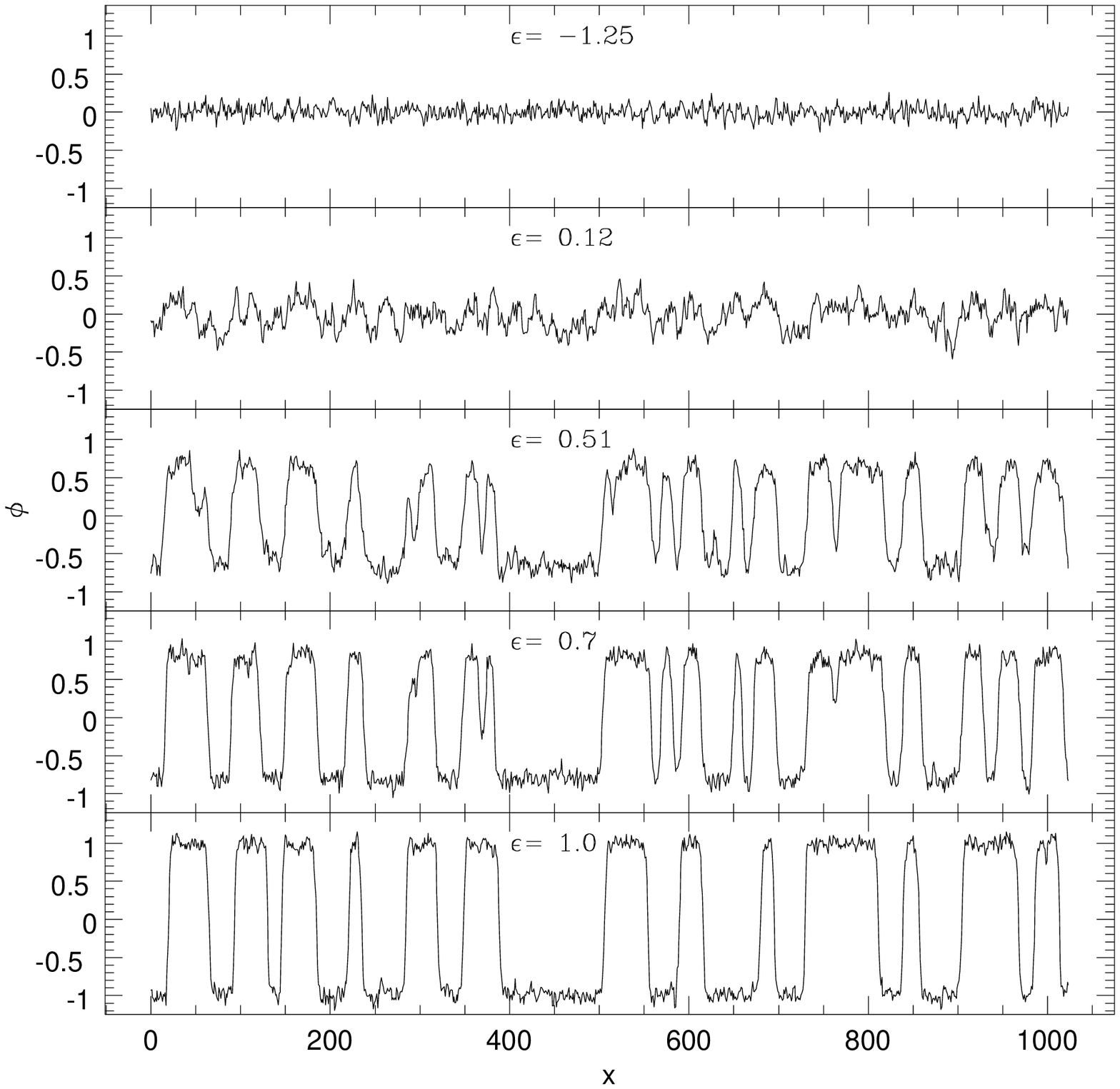}
\caption[figure2]{\label{figure2}
Snapshots of $\varphi$ during 
kink formation with a quench timescale of $\tau_Q= 64$ and damping
parameter $\eta = 1$.
The figures, from top to bottom, correspond to $ t  = -80.0, 7.5, 32.5,
45.0$ and $333.0$,
respectively. 
}
\end{figure}

We investigate the creation of kinks as a function
of $\tau_Q$ by starting at some $\epsilon < 0$ suitably above
the transition, and then gradually adjusting the shape of the potential
in accord with Eq.~(\ref{eq:epsilon}). 
Figure\ \ref{figure2} shows a sequence of
``snapshots'' of $\varphi$ obtained in the course of such a quench. When
$\epsilon < 0$, $\varphi$ fluctuates around its expectation
value $\langle\varphi\rangle = 0$. 
The same situation initially persists for slightly positive  $\epsilon$.
However, further below $T_C$, $\varphi$
settles locally around one of the two alternatives:
$ \langle\varphi\rangle = \pm \sqrt { \epsilon } \ . $
Moreover, local choices of one of these two 
alternatives cannot be easily undone once a certain symmetry breaking 
is selected, unless the kinks are separated
by distances no larger than $\xi$ (see Fig.\ \ref{figure2}).

To test the predictions of defect density in Ref. \cite{Zurek85}, we note again that,
for a sizable $\eta$ and sufficiently small $\epsilon$, the 
damping term $\eta \dot \varphi$ in Eq.~(\ref{eq:motion}) is bound to dominate. 
Such overdamped evolution will take place whenever
$ \eta^2 > | \epsilon | \ $\cite{Antunes96}.
In our case, $\eta > 1$ and $ | \hat \epsilon | \ll 1$.
The characteristic relaxation time is then given by
$ \tau = \eta / |\epsilon| \ . $
Consequently, $\mu = 1$ in Eq.~(\ref{eq:tau}), and
\be
\hat t = \sqrt{ \eta \, \tau_Q} \ . 
\label{eq:cond1}
\ee
The corresponding value of the relative temperature immediately yields,
$
\hat \epsilon = \hat t / \tau_Q = \sqrt{\eta / \tau_Q} \ . 
$
It follows that
\be
\hat \xi = \xi_0 ~(\tau_Q/\eta)^{1/4} \ ,
\label{eq:cond3}
\ee 
where we have adopted $\nu = 1/2$ in accord with the Landau-Ginzburg theory
and in agreement with the critical exponents inferred 
from the behavior of the healing length in Fig.\ \ref{figure1}. 
Eqs. (\ref{eq:cond1}) and (\ref{eq:cond3}) 
are expected to be applicable as long as the condition $ \eta^2 > | \epsilon |$ holds 
at $\hat \epsilon$, which in turn implies $\eta / \tau_Q > 1$. 

\begin{figure}[h]
\leavevmode
\epsfxsize=3.2truein\epsfbox{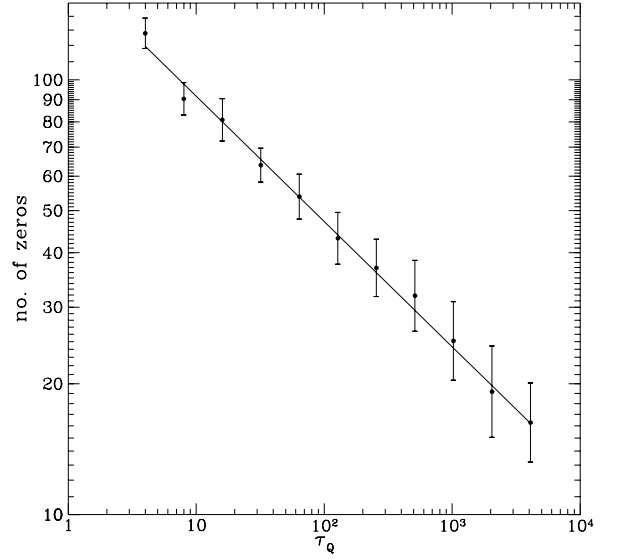}
\caption[figure3]{\label{figure3}
Number of defects as a function of quench timescale.
The straight line is the best fit to $ n \propto \tau_Q^{-a}$
with $a = 0.28 \pm 0.02$ ($\chi^2 = 1.96$).  This 
exponent compares favorably with the theoretical prediction of $1/4$ based on 
the theory in Ref. \cite{Zurek85}.
}
\end{figure}

Figure\ \ref{figure3} is the principal result of our paper. It illustrates the density 
of kinks obtained in a sequence of quenches with $\eta = 1$ for various
$\tau_Q$ values.
For each $\tau_Q$,
the phase transition was simulated 15 times, starting at
$\epsilon = -1$ (except for the shortest and longest $\tau_Q$,
which were initiated at $\epsilon = -10/\sqrt{\tau_Q}$).
Our computational domain had periodic boundary conditions.
Production runs were carried out with resolution of 16,384 
gridpoints. Convergence was checked by comparing results at different resolutions.
The ``physical'' size of the computational ring in the production runs was 2,048 
units. At this scale, the ring is large enough, so boundary
effects are avoided.

\begin{figure}[h]
\leavevmode
\epsfxsize=3.2truein\epsfbox{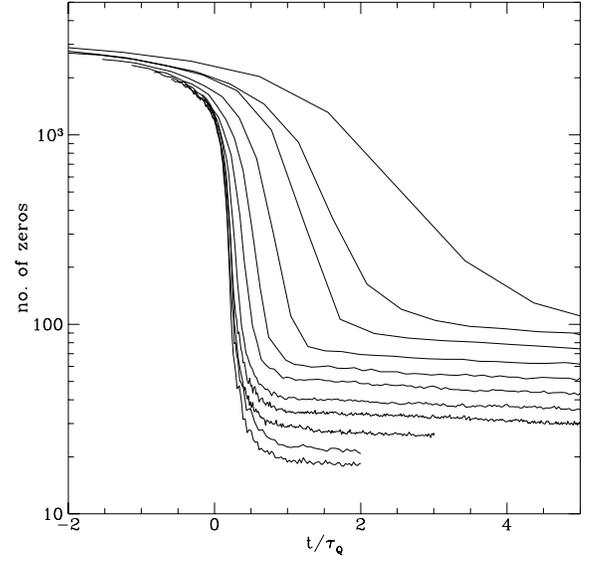}
\caption[figure4]{\label{figure4}
Average number of zeros as a function of time in units of $\tau_Q$;
from top to bottom $\tau_Q = 4, 8, ..., 2048, 4096$. 
The number of kinks used in Fig.\ \ref{figure3} were 
obtained at $t/\tau_Q \sim 4$, except in the large, computationally expensive,
$\tau_Q$ cases where an extrapolated value was used.
}
\end{figure}

The number of kinks produced by the quench was obtained by counting 
the number of zeros of $\varphi$. Above and immediately below $T_C$,
there is a significant number of zeros which have 
little to do with the kinks (see Fig.\ \ref{figure2}). 
However, as the quench proceeds, the number of zeros quickly evolves towards
an ``asymptotic'' value 
(see Fig.\ \ref{figure4}). This change of the density of zeros and 
its eventual stabilization is associated with 
the obvious change of the character of $\varphi$ and with 
the appearance of the clearly defined kinks.
By then, the number of kinks is nearly constant in the runs with long
$\tau_Q$, and, correspondingly, their kink density is low. Even in the runs
with the smallest $\tau_Q$, there is still a clear break 
between the post-quench rates of the disappearance of zeros and the long-time,
relatively small rate at which the kinks annihilate. 

In the regime here investigated, the theoretical scaling relation 
for the number density of kinks,
$ n \approx (\eta/\tau_Q)^{1 \over 4} \approx 1/f\,\hat\xi \ , $
appears to be well satisfied, with $f \sim 8$.
It is interesting to note that a similar value of $f$ has been
deduced from the experiment in Ref. \cite{Bauerle}. 
Experimentally we find 
$n = (8.7\times 10^{-2} \pm 1.0\times 10^{-3}) \tau_Q^{-0.28 \pm 0.02}$,
when the kinks are counted at approximately the same $t/\tau_Q$ value.
A similar scaling is also obtained for kinks counted at equal $t$ times
after the quench.

In summary, our numerical experiment appears to provide a strong confirmation of the
theoretical predictions given by one of us \cite{Zurek85}.
As expected, the density of topological defects is somewhat 
less than the inverse of $\hat \xi$ but of the right order of magnitude. 
Most importantly, the scaling of the kink density with $\tau_Q$ follows closely theoretical
expectations. These results are also supported by the dependence
of the number of kinks on the damping parameter $\eta$, as well as by the
preliminary results of the computer experiments involving complex order
parameter and/or more than one dimensional space \cite{LagunaZurek96}.

We thank S. Habib for helpful discussions.
This work was partially supported by 
NSF grants PHY 93-09834, 93-57219 (NYI) to P.L. and
NASA HPCC to W.H.Z.

%
%

%
%

%
%

%
%

\end{document}